\documentclass[10pt]{iopart}
\usepackage{iopams}
\usepackage{cite}
\usepackage{graphicx,color}% Include figure files
\usepackage{dcolumn}% Align table columns on decimal point
\usepackage{bm}% bold math

\begin{document}

\title{Monovacancy paramagnetism in neutron-irradiated graphite probed by $^{13}$C NMR}

\author{Z T Zhang$^{1,2}$, C Xu$^{1}$, D Dmytriieva$^{2,3}$ S Molatta$^{2,3}$, J Wosnitza$^{2,3}$, Y T Wang$^{1,4}$, M Helm$^{1,3}$, Shengqiang Zhou$^{1}$ and H K\"{u}hne$^{2}$}

\address{$^{1}$ Institute of Ion Beam Physics and Materials Research, Helmholtz-Zentrum Dresden-Rossendorf, D-01314 Dresden, Germany}
\address{$^{2}$ Hochfeld-Magnetlabor Dresden (HLD-EMFL), Helmholtz-Zentrum Dresden-Rossendorf, D-01314 Dresden, Germany}
\address{$^{3}$ TU Dresden, D-01062 Dresden, Germany}
\address{$^{4}$ School of Microelectronics, Key Laboratory of Wide Band-Gap Semiconductor Materials and Devices, Xidian University, Xi'an 710071, China}
%\address{$^{6}$ Author to whom any correspondence should be addressed.}

\eads{\mailto{z.zhang@hzdr.de} and \mailto{h.kuehne@hzdr.de}}
%\ead{}

\begin{abstract}
We report on the magnetic properties of monovacancy defects in neutron-irradiated graphite, probed by $^{13}$C nuclear magnetic resonance spectroscopy. The bulk paramagnetism of the defect moments is revealed by the temperature dependence of the NMR frequency shift and spectral linewidth, both of which follow a Curie behavior, in agreement with measurements of the macroscopic magnetization. Compared to pristine graphite, the fluctuating hyperfine fields generated by the defect moments lead to an enhancement of the $^{13}$C nuclear spin-lattice relaxation rate $1/T_{1}$ by about two orders of magnitude. With an applied magnetic field of 7.1 T, the temperature dependence of $1/T_{1}$ below about 10 K can well be described by a thermally activated form, $1/T_{1}\propto\exp(-\Delta/k_{B}T)$, yielding a singular Zeeman energy of ($0.41\pm0.01$) meV, in excellent agreement with the sole presence of polarized, non-interacting defect moments.
\end{abstract}

%\pacs{76.60.-k, 75.50.Pp, 61.72.J-}
\vspace{2pc}

\noindent{(Some figures may appear in colour only in the online journal)}

\submitto{\JPCM}
\maketitle
\ioptwocol

\section{Introduction}
The magnetism of carbon-based materials has attracted a lot of interest during the past decade \cite{rev-nano-carbon,rev-graphene-spintronics,rev-RepProPhys,Science-painting,Science-control-H,Nat-RT-zig-ribbon,Nc-graphene-switchByDoping,SiC-qubit1,SiC-qubit2,phosphorene,NJP-switch-momment,NJP-substi-DFT,NJP-DFT-nanotube,graphite-pi-electron,PRL-graphene-STM,PRL-RT-H-Graphene,Np-spin-half,PRL-graphene-spinCurr,PRL-Missing-Atom,NJP-Ohldag-graphite-H,Np-RT-graphite-ptDfect,PRL-proton-graphite,graphite-C-imp,graphite-H-He}.
As one of the main reasons, it offers a new conceptual framework to study the unconventional magnetism of the $sp$-electrons \cite{PRL-graphene-STM,PRL-RT-H-Graphene,Np-spin-half,PRL-graphene-spinCurr,PRL-Missing-Atom,NJP-Ohldag-graphite-H,Np-RT-graphite-ptDfect,graphite-pi-electron,PRL-proton-graphite,graphite-C-imp,graphite-H-He,DFT-moment-vanish,NJP-substi-DFT,DFT-Kondo-pi-nonVanish,DFT-PRB-sigma-pi,NJP-DFT-RTFM}. The theoretical modeling of the local-moment formation in the related materials is still controversial \cite{DFT-moment-vanish,NJP-substi-DFT,DFT-Kondo-pi-nonVanish,DFT-PRB-sigma-pi,NJP-DFT-RTFM}. For example, a density functional theory (DFT) study by Palacios \emph{et al.} showed that the vacancy-induced $\pi$ moments in graphene quench at any experimentally relevant vacancy density \cite{DFT-moment-vanish}, whereas Nanda \emph{et al.} suggested that both the $\sigma$ and $\pi$ electrons of the vacancy defects contribute and favor an $S=1$ state with a reduced net magnetic moment of 1.7 $\mu_{\mathrm{B}}$ \cite{NJP-substi-DFT}. Experimentally, there is emerging consensus on magnetism in carbon-based systems. Evidence of $\pi$ magnetism is, for example, provided by X-ray dichroism and scanning tunneling microscope techniques \cite{graphite-pi-electron,PRL-graphene-STM}. However, both itinerant $\pi$ electrons and dangling $\sigma$ bonds are proposed to account for the vacancy-induced paramagnetism of spin-$1/2$ moments in graphene \cite{Nc-graphene-switchByDoping}.
Both in graphene and graphite, monovacancy defects are theoretically proposed and experimentally identified as a primary source of defect-generated magnetism \cite{planarity-ground-state,NJP-substi-DFT,DFT-PRB-sigma-pi,DFT-Kondo-pi-nonVanish,coulomb-charging,H-complexs,DFT-3D-graphitic,Yazyev-graphite-stacking,PRL-graphene-STM,PRL-Missing-Atom}. As suggested by DFT studies, the local moment of the vacancies can range from 0.06~$\mu_{B}$ to 1.7~$\mu_{B}$, depending on factors such as, for example, interactions between the (quasi-)localized vacancy-$\sigma$/$\pi$ states and the itinerant Dirac states, lattice deformations (or nonplanarity), or the defect concentration \cite{NJP-substi-DFT,planarity-ground-state,graphene-defect-density,DFT-3D-graphitic}.

Since the dilute bulk distribution of the magnetic defects leads to an only very weak macroscopic magnetization, conventional magnetometry studies inherently suffer from the presence of magnetic impurities, contaminations, and background problems, all of which are complicating the investigation of  the intrinsic defect properties \cite{purity-substrate1,EPL-CommentPristine-graphite,purity-substrate3,purity-substrate4}. These problems can be avoided by using nuclear magnetic resonance (NMR) as a local probe for detecting the intrinsic magnetic properties. NMR measurements are rarely performed in the investigation of defect magnetism \cite{graphite-NMR-srep}, mainly due to the limited instrumental sensitivity that prevents the study of samples with a very small total number of nuclear spins, such as monolayer graphene and ion-irradiated thin films of graphite.

Recently, some of us introduced bulk paramagnetism into graphite by intentionally creating defects through neutron irradiation \cite{YTWang-graphite}.
Here, the irradiation leads to the local displacement of one carbon atom. Each of the three remaining, neighboring atoms is left with one $sp^2$ dangling bond, two of which are then coupled to leave the third $sp^2$ bond unpaired, thus generating a localized $S = 1/2$ moment. Transplanar defects are likewise considered as a source for localized magnetism.

In this paper, we study the intrinsic magnetism in graphite with an average defect concentration of about 1.24$\times 10^{-3}$ per formula unit by means of $^{13}$C NMR spectroscopy. The bulk paramagnetism of the defects is reflected by the temperature dependence of the NMR frequency shift and spectral linewidth, both of which follow a Curie behavior. In the dynamical properties, the nuclear spin-lattice relaxation rate $1/T_{1}$ is, due to the presence of the fluctuating defect moments, significantly increased with respect to pristine graphite. With a magnetic field of 7.1 T applied out-of-plane, $1/T_{1}$ follows a thermally activated behavior $\propto$ $\exp(-\Delta/k_{B}T)$ with a singular gap energy of $\Delta=(0.41\pm0.01)$ meV, in excellent agreement with the Zeeman splitting of non-interacting $S = 1/2$ defect moments.

\section{Experiment}
Samples of highly oriented pyrolytic graphite (HOPG) with grade ZYA were used, and are generally referred to as graphite in this paper. The neutron irradiation was performed at the reactor BER II (Position DBVK) at Helmholtz-Zentrum Berlin \cite{Neutron-exp-detial1}. During irradiation, the temperature of the samples was less than 50 $^{\circ}$C \cite{Neutron-exp-detial2}. The sample used for our NMR measurements was irradiated for 150 hours with the fluence reaching $3.12\times10^{19}$ cm$^{-2}$ (only fast neutrons) \cite{YTWang-graphite}.

The $^{13}$C NMR spectra and spin-lattice relaxation times $T_{1}$ were recorded via Hahn spin-echo detection with a magnetic field of 7.100 T applied in $H\parallel{c}$ (out-of-plane) direction. For a given temperature, $T_{1}$ was obtained from fitting
\begin{eqnarray}\label{T1Fit}
M_{z}(\tau)=M_{0}\{1-\exp[-(\frac{\tau}{T_{1}})^{\lambda}]\}
\end{eqnarray}
to the recovery of the nuclear magnetization after saturation, where $\lambda$ is introduced as a stretching exponent to account for a distribution of relaxation times accross a given NMR spectrum. The magnetization was measured using a superconducting quantum interference device-vibrating sample magnetometer (SQUID-VSM). All of the experiments were carried out on the same neutron-irradiated graphite (NI graphite) sample.
For means of comparison, $^{13}$C NMR spectra and  $T_{1}$ of pristine graphite
were also recorded at selected temperatures.

\section{Results and discussion}

%\subsection{$^{13}$C NMR spectra - static properties}

The $^{13}$C NMR spectra correspond to the transition $I_z = -1/2 \rightarrow +1/2$ between the Zeeman-split nuclear Eigenstates. The relative NMR frequency shift $K$ is defined as $K=(f_{n}-f_{0})/f_{0}$, where $f_{0}={\gamma}\mu_{0}H/2\pi$ is the Larmor frequency of a bare nucleus with a gyromagnetic ratio ${\gamma}$ in a magnetic field $\mu_{0}H$, and $f_{n}$ is the observed NMR frequency. In the present case, $K$ is the sum of local fields coupled to both spin and orbital moments: $K=K_{spin}+K_{orb}$. The spin part $K_{spin}=A_{hf}\chi_{spin}$ is the product of the electronic uniform spin susceptibility $\chi_{spin}$ and the hyperfine coupling $A_{hf}$. The orbital part $K_{orb}$ is temperature independent and gives the main contribution to the NMR shift at high temperatures, whereas $K_{spin}$, probing the dipole fields at the nuclear sites stemming from the defect moments, is of paramagnetic nature and decays rapidly with increasing temperature.

\begin{figure}[tbp]
	\centering
	\includegraphics[width=8.0cm]{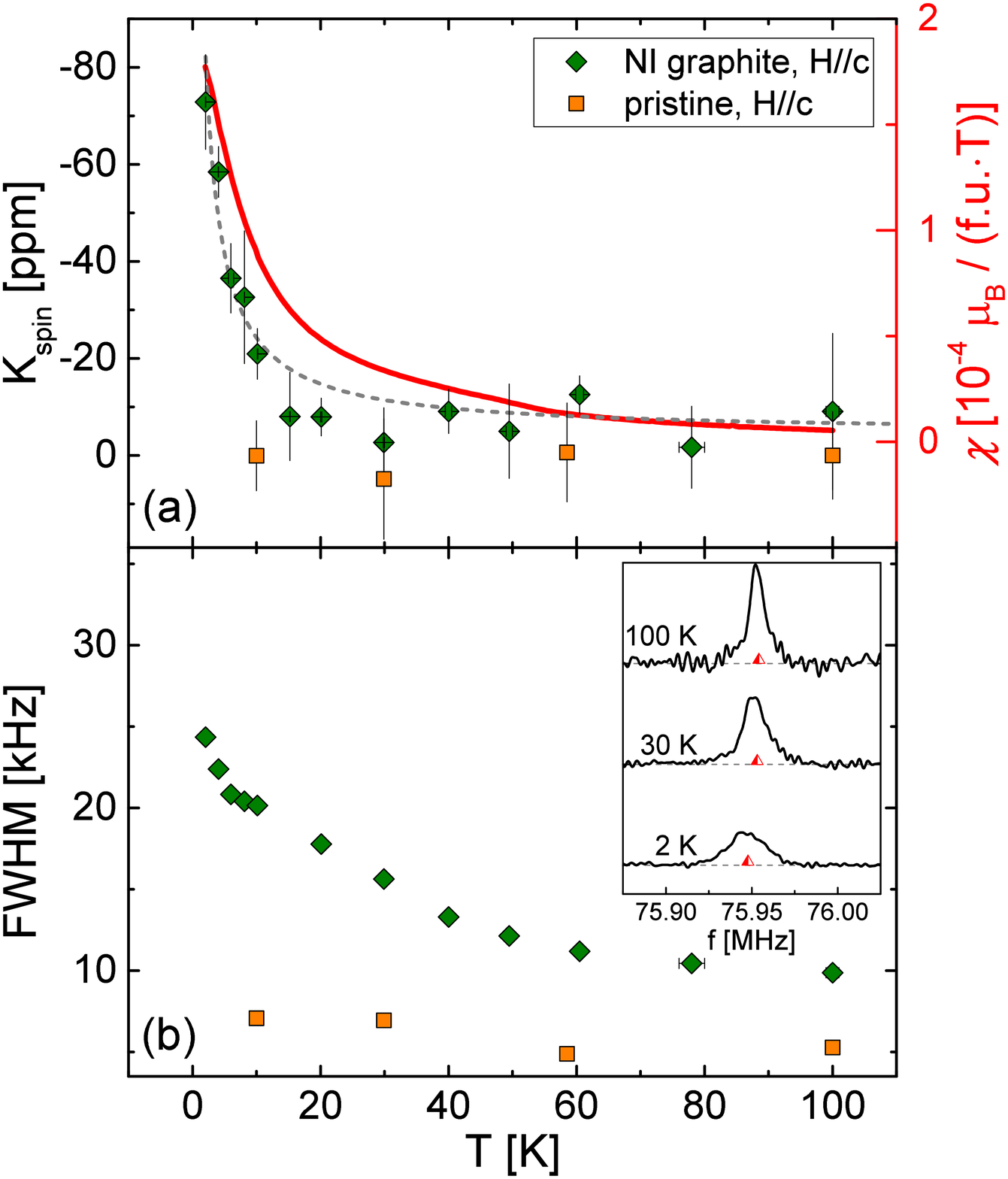}
	\caption{\label{Fig_T_shift_FWHM} (a) Temperature dependence of the shift $K_{spin}$, probing the average local spin susceptibility. The red line represents the macroscopic DC susceptibility of NI graphite for an out-of-plane applied field of 7 T. A Curie-Weiss fit to $K_{spin}$ (see text) is indicated by the gray, dashed line. (b) Temperature-dependent $^{13}$C NMR linewidth, reflecting the width of the internal dipole-field distribution. The inset shows three spectra at selected temperatures, each with the first spectral moment indicated by a red triangle.}
\end{figure}

Therefore, $K_{spin}$ can be extracted in good approximation as $K(T)-K(\mathrm{100~K})$. Figure~\ref{Fig_T_shift_FWHM}(a) shows the obtained values of $K_{spin}$ for both NI and pristine graphite as a function of temperature in comparison with the macroscopic susceptibility $\chi(T)$ of NI graphite. For the latter, the scaling factor between $K_{spin}$ and $\chi$ gives an average hyperfine coupling constant $A_{hf}$ $\approx$ -0.36$\pm$0.03 f.u.$\times$T/$\mu_{\mathrm{B}}$, or $\approx$ -4.5$\pm$0.3$\times10^{-4}$ T/$\mu_{\mathrm{B}}$, respectively. This comparably small value is in agreement with dipole-dipole interactions between the vacancy and nuclear moments over an average distance of several nanometers. With the DC susceptibility  almost fully saturated at 2 K and $\mu_0 H =$ 7 T, compare Fig. 1 (a), and assuming 1 $\mu_B$ per monovacancy defect, we evaluate the average defect
concentration as 1.24$\times 10^{-3}$ per formula unit.

The temperature dependence of $K_{spin}$ shows a typical paramagnetic behavior for NI graphite, in very good agreement with the macroscopic DC susceptibility $\chi(T)$, the latter being in line with our previously reported results \cite{YTWang-graphite}. In comparison, $K_{spin}$ of pristine graphite appears as negligible, so that the increased shift in NI graphite can clearly be attributed to the defect magnetism. Furthermore, the observed temperature dependence of $K_{spin}$ points to an anisotropic spatial distribution of the defect moments. For neutron-irradiated SiC (NI SiC), we have addressed this issue by modeling the microscopic distribution of dipole fields generated by the defect moments \cite{my-SiC}. From a Curie-Weiss fit to the NMR shift according to $K_{spin} = K_{spin,0} + A_{hf} \times C/(T-\Theta) $, where $K_{spin,0}= -5$ ppm accounts for a small uncertainty in approximating $K(\mathrm{100~K})$ as the high-temperature limit, we find a Curie-Weiss temperature $\Theta =$ (-0.7 $\pm 1.0$) K. Therefore, any interactions between the defect moments would be very small compared to our experimental temperature range.

The intrinsic paramagnetism in NI graphite is further confirmed by the Curie behavior of the temperature-dependent linewidth, compare Fig.~\ref{Fig_T_shift_FWHM}(b), which quantitatively reflects the width of the local dipole-field distribution, scaling with the paramagnetic amplitude of defect moments. Again, the comparison with pristine graphite clarifies the contribution of the intrinsic paramagnetism to the linewidth.  At 100 K, the linewidth of NI graphite is about twice as large as for pristine graphite. We attribute this observation to an increased distribution of local orbital moments, caused by irradiation-induced strain and displacement effects in the lattice.

%\subsection{Nuclear spin-lattice relaxation rate}

%Next, we discuss the fluctuations of the defect moments and the related dynamic dipole fields at the $^{13}$C nuclear positions, leading to an increase of the spin-lattice relaxation rate compared to pristine graphite.

Next, we discuss the coupling of the nuclear 
spin-lattice relaxation rate to the dynamical properties of the defect moments. In general, the $T_{1}$ relaxation is driven by the transverse components of the field fluctuations $\delta \mathbf{h}(t) = \mathbf{h}(t) - \left\langle \mathbf{h}(t) \right\rangle $ at the nuclear site \cite{Horvatic2002}:
\begin{eqnarray}\label{T1hypfluct}
\frac{1}{T_1}= \frac{\gamma_n^2}{2} \sum_{\alpha=x,y} \int_{-\infty}^{+\infty}{
	 \left\langle \delta h_{\alpha}(t) \delta h_{\alpha}(0) \right\rangle ~e^{-i \omega_n t}~dt},
\end{eqnarray}
where $\omega_n$ is the nuclear Larmor frequency. In NI graphite, $\delta \mathbf{h}(t)$ stems mainly from the fluctuations of the nearest vacancy moments, which are coupled to a given nuclear moment via dipole--dipole interaction such that $\delta \mathbf{h}(t) = - \stackrel{\leftrightarrow}{\mathbf{A}} \cdot \mathbf{S} (t)$. The dipolar coupling tensor is defined as
\begin{equation}\label{diploe_field}
\stackrel{\leftrightarrow}{\mathbf{A}}=\frac{\mu_{0}}{4\pi}\sum_{j}\left(\frac{3\mathbf{r}_{j}\mathbf{r}_{j}}{r^{5}_{j}}-\stackrel{\leftrightarrow}{\mathbf{\delta}} \frac{1}{r^{3}_{j}}\right).
\end{equation}
Here, $\mu_{0}$ denotes the vacuum permeability, and $\mathbf{r}_{j}$ is the position vector of the nucleus relative to the $j$-th defect moment,
with a relevant radius of contributing defects of a few ten nanometers typically. 
%The local field fluctuations can now be written as
%\begin{eqnarray}
%\delta h_{\alpha}(t)=\sum_{\beta}A_{\alpha \beta} S_\beta (t),
%\end{eqnarray}
%where $\alpha$ and $\beta$ represent the crystallographic axes $(x,y,z)$. 

Assuming a magnetic field along the $z$ direction, $1/T_1$ can be expressed in terms of the vacancy moment autocorrelation function $\langle S_{\beta} (t) S_{\beta} (0) \rangle$ (with $\beta = (x,y,z))$:
\begin{eqnarray}\label{T1fluct}
\frac{1}{T_1} &=& \frac{\gamma_n^2}{2} \int_{-\infty}^{+\infty} ( [A_{x z}^2 + A _{y z}^2] \langle  S_{z} (t)  S_{z} (0) \rangle \nonumber \\ &+& [A_{x x}^2 + A_{y x}^2] \langle  S_{x} (t)  S_{x} (0) \rangle  \\ &+& [A_{x y}^2 + A _{y y}^2] \langle  S_{y} (t)  S_{y} (0) \rangle ) e^{-i \omega_n t} ~dt.\nonumber
\end{eqnarray}
With the given random spatial distribution of monovacancy moments, the dipole fields are locally well defined, but essentially randomly distributed over the sample volume, similar to the results of the simulations that we have perfomed in our investigation of the defect magnetism in NI SiC \cite{my-SiC}.

Hence, the hyperfine coupling tensor varies locally, leading to a distribution of scaling factors $A_{\alpha \beta}$ to the autocorrelation function in Eq. (\ref{T1fluct}). In consequence, $T_1$ is a singular value for a given nuclear site, and approaches a finite-width distribution of values when integrating over a large real-space volume with many nuclear spins. Correspondingly, the stretching exponent $\lambda$, introduced in Eq. (\ref{T1Fit}), is smaller than unity. On the other hand, temperature-dependent changes to the autocorrelation function result in the same relative variation of $T_1$ for all nuclear spins, but leave $\lambda$ unchanged. In consequence, when the fluctuations of the vacancy moments are gapped, the temperature-dependent relaxation of all nuclear spins will exhibit the same Arrhenius-type activated behavior, yielding a singular gap value.

\begin{figure}[tbp]
	\centering
	\includegraphics[width=8.0cm]{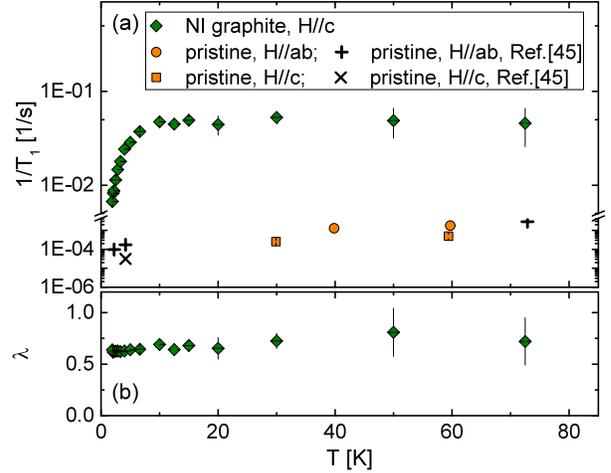}
	\caption{\label{Fig_T_1T1T} (a) Temperature dependence of $1/T_{1}$ for NI and pristine graphite. Some of the data for pristine graphite (``$+$'' and ``$\times$'') are from Ref. \cite{NMR-pure-graphite}. (b) Temperature-dependent stretching exponent $\lambda$ of the $T_{1}$ relaxation for NI graphite.}
\end{figure}

Figure~\ref{Fig_T_1T1T}(a) shows the temperature-dependent spin-lattice relaxation rate for the NI graphite sample in comparison to pristine graphite. In case of the pristine sample, the $1/T_{1}$ relaxation rate, driven only by weak itinerant fluctuations, is very small and increases slightly with increasing temperature, in agreement with previously reported results by Maniwa et al. \cite{NMR-pure-graphite}. For the NI graphite sample, $1/T_{1}$ is enhanced by about two orders of magnitude. As shown in Figure~\ref{Fig_T_1T1T}(b), the stretching exponent $\lambda$ ranges from 0.57 to 0.85 between 2 and 100 K, in line with a distribution of $T_{1}$ values across the spectral line as discussed above. In contrast, a spatially homogeneous susceptibility generated by itinerant electrons would result in a stretching exponent close to unity. Therefore, the observed distribution of $T_{1}$ values clearly indicates that the spin-lattice relaxation mechanism is dominantly caused by fluctuations of localized defect moments. Above about 10 K, $1/T_{1}$ appears as temperature independent, which is in line with paramagnetic fluctuations when $k_{\mathrm{B}}T$ is large compared to the Zeeman energy. Below 10 K, however, a strongly decreasing trend of $1/T_{1}$ is observed, indicating a gapped behavior of the spin fluctuations \cite{Moriya-paramagnetic-salt,Moriya-paramagnetic-expo}.

In general, both the electronic spin-lattice and spin-spin relaxation mechanisms may determine the vacancy moment correlations introduced above. Since the defect moments are coupled to all surrounding $^{13}$C nuclear moments in a volume of several ten cubic nanometers, the low-temperature spin-spin relaxation can well be expected as much faster than the electronic spin-lattice relaxation \cite{Tyryshkin2011}.
Therefore, the nuclear spin-lattice relaxation at low temperatures is mostly determined by the longitudinal component of the vacancy moments. One- and two-phonon scattering are known to drive the electronic spin-lattice relaxation in many solid-state materials. Two-phonon scattering of the Orbach type may determine the electronic spin-lattice relaxation $\propto \exp(-\Delta/k_{B}T)$ under the presence of an excitation gap $\Delta$ \cite{Orbach1972}. Presumably, this process also drives the low-temperature nuclear-spin lattice relaxation in graphite.

At high temperatures, where $k_{B}T$ is much larger than all electronic and nuclear interaction energies,  the regime of fast electronic motions is approached, and $1/T_{1}$, probing the spectral density at the nuclear Larmor frequency, is constant. This is a commonly observed phenomenology in the paramagnetic limit of materials with strongly localized materials \cite{Kuehne2009,Utz2015,Melzi2001}

\begin{figure}[tbp]
\centering
\includegraphics[width=8.0cm]{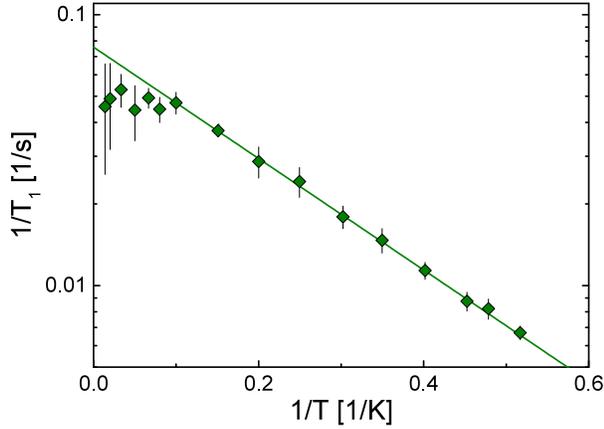}
\caption{\label{Fig_Thermal_active_T1} $1/T_{1}$ in logarithmic scale as a function of reciprocal temperature for the NI graphite sample with $H\parallel{c}=7.1$ T. The solid line indicates a fit according to $1/T_{1} \propto \exp(-\Delta/k_{B}T)$.}
\end{figure}

For further investigation of the gapped behavior, the $1/T_{1}$ data is plotted in logarithmic scale as a function of the reciprocal temperature $1/T$ in Fig.~\ref{Fig_Thermal_active_T1}. Below 10 K, $1/T_{1}$ can very well be described by a thermally activated behavior according to $1/T_{1} \propto \exp(-\Delta/k_{B}T)$, with an excitation gap of $\Delta=(0.41\pm0.01)$ meV.
As for the origin of this gap, dipole--dipole interactions between the defect moments can be ruled out. Firstly, the defects created by neutron irradiation are, apart from a certain anisotropy that leads to a finite shift $K$, randomly positioned in the sample. Any gapped behavior based on dipole-dipole interactions, which are orientation and distance dependent, would exhibit a broad distribution of gap values, in stark contrast to our observations. Secondly, the typical distance between two paramagnetic centers at this defect concentration is estimated as several nanometers \cite{my-SiC}. The dipole-dipole interaction between two defect moments of $1~\mu_{B}$ over this distance is much less than 1 $\mu$eV, which is negligible compared to the observed gap magnitude.

Rather, we consider the Zeeman splitting of defect states as the origin of the observed excitation gap, depending only on the spin moment and the applied magnetic field, but not on the spatial distribution of the noninteracting moments. As established by magnetometry and ESR studies of the vacancy moments in neutron- and proton-irradiated graphite, respectively, the magnetic monovancy states are associated with a spin moment $S=1/2$, a magnetic moment of $1~\mu_{B}$ and an electronic g-factor of about 2 \cite{YTWang-graphite,Lee2006}. Correspondingly, the Zeeman energy in an applied field of $B = 7.1$ Tesla is $ g\mu_B  m_z B = 0.41$ meV (or 4.76 K), which exactly equals the observed excitation gap. Since there is no significant change of the stretching exponent down to lowest temperatures, we can conclude that the gap is of a singular value, corroborating the absence of a notable interaction between the vacancy moments. Otherwise, a finite coupling of defect moments would lead to a modification of the local excitation gaps, and, in the present case of randomly positioned defects, result in a finite-width distribution of gap values with corresponding changes to the stretching exponent at temperatures of the order of the average gap energy.

%. In that case, however, an increased broadening of the $T_{1}$ distribution would be reflected by a temperature-dependent stretching exponent at temperatures of the order of the respective gap values.

%\tr{Probably discuss similar as in SiC that ferromagnetism would not be seen by NMR because of small volume fraction or fast depolarization or because it does not occur for the given defect concentration.}

\section{Summary}
In summary, we have investigated the intrinsic magnetic properties of monovacancy defects in neutron-irradiated graphite by means of $^{13}$C NMR spectroscopy. The paramagnetic nature of the defect-induced magnetism is revealed by the temperature dependence of the NMR shift and spectral linewidth, both of which follow a Curie behavior, with negligible Curie-Weiss temperature within experimental error. The nuclear spin-lattice relaxation rate $1/T_{1}$ is driven mostly by fluctuations of the localized defect moments rather than by itinerant susceptibility.  With a magnetic field of 7.1 T applied out-of-plane, the temperature dependence of $1/T_{1}$ below about 10 K can well be described by a thermally activated behavior, which we conclude to stem solely from the Zeeman interaction of the noninteracting vacancy moments with the external magnetic field. Our findings clarify the intrinsic magnetic properties of the vacancy moments in neutron-irradiated graphite and, thus, provide fundamental information for the theoretical modeling of defect-induced magnetism in carbon-based materials.

\ack
The neutron irradiation was done at the Helmholtz-Zentrum Berlin f\"{u}r Materialien und Energie by Gregor Bukalis. The project is supported by Helmholtz-Association (VH-PD-146). Z.T.Z. was financially supported by the National Natural Science Foundation of China (Grant No. 11304321) and by the International Postdoctoral Exchange Fellowship Program 2013 (Grant No. 20130025). Further, support by the HLD at HZDR, a member of the European Magnetic Field Laboratory, and by the Deutsche Forschungsgemeinschaft (DFG) through the Research Training Group GRK 1621 is gratefully acknowledged.

\section*{References}
\bibliography{References}

%\section{to-do}
%
%- Nochmal Formeln T1 Mechanismus durchgehen\\

\end{document}